\documentclass[11pt]{article}
\usepackage[ascii]{inputenc}  
\usepackage[protrusion=false]{microtype}
\usepackage{amsmath}
\newtheorem{definition}{Definition}[section]
\newtheorem{claim}[definition]{Claim}
\newcommand{\var}[1]{\mathit{#1}}
\title{
  Critique of Boyu Sima's Proof that P $\neq$ NP}
\author{Brendon Pon\\
Department of Computer Science\\
University of Rochester\\
Rochester, NY 14627, USA}
\date{May 7, 2020}
\showhyphens{}
\newcommand{\np}{{\rm NP}}
\newcommand{\p}{{\rm P}}
\newcommand{\pisnp}{\ensuremath{\p=\np}}
\newcommand{\pisnotnp}{\ensuremath{\p\neq\np}}

\begin{document}

\maketitle

\begin{abstract}
  We review and critique Boyu Sima's paper, ``A solution of the {P}
  versus {NP} problem based on specific property of clique function,''
  which claims to prove that $\pisnotnp$ by way of removing the
  gap between the nonmonotone circuit complexity and the monotone
  circuit complexity of the clique function. We first describe Sima's
  argument, and then we describe where and why it fails. Finally, we present
  a simple example that clearly demonstrates the failure.
\end{abstract}

\section{Introduction}
One of the most well-known and long-standing problems in computer science is the question of whether $\pisnp$. A solution to the problem would have wide-ranging implications to everything from economics to cybersecurity. To this end many have claimed to have found a proof that either $\pisnp$ or $\pisnotnp$. However, to this date no such claim has been found to be correct.
 
 There are various methods for attempting such a proof. One such
 method is by using lower bounds on the complexity of circuits. By
 showing that a known NP-complete problem has an exponential
 lower-bounded circuit complexity, you show that $\pisnotnp$. In his
 paper ``A solution of the P versus NP problem based on specific
 property of clique function''~\cite{sima2019}, Sima tries to do
 precisely this. Sima analyzes the clique function and attempts to
 show that the circuit complexity of the clique function is
 exponential, thus showing that $\pisnotnp$.
 
 In this paper, we will first present some definitions and some prior work that Sima uses in his argument. We will then present Sima's argument and describe where Sima's argument fails due to making an improper generalization and failing to consider the connection between a Boolean variable and its negation. Finally we will provide an example that demonstrates the hole in his algorithm.  
\section{Preliminaries}
The following are the needed definitions and an existing theorem that
will be used.
 \subsubsection*{Boolean Functions}
 A Boolean function of $k$ variables is a function $f: \{0, 1\}^k \rightarrow \{0,1\}$.
 
Boolean functions (of $k$ variables) can be expressed as Boolean (propositional) formulas with $k$ variables and the logical operators ($\wedge, \lor, \lnot$). 

A Boolean function $f$ of $k$ variables is called monotone if 
\begin{equation*}
    (\forall w, w'\in \{0,1\}^k: w\leq_{\mathrm lex} w') [f(w)\leq f(w')]. 
\end{equation*}

\noindent Note: If a function is monotone, then changing a 0 to a 1 in the input will never cause a decrease in the output, and changing a 1 to a 0 in the input will never cause an increase in the output.
 \subsubsection*{Boolean Circuits}
 A Boolean circuit is a directed acyclic graph with \emph{gate} nodes and \emph{input} nodes. Gate nodes can be one of three types corresponding to the logical operators AND ($\wedge$), OR ($\lor$), and NOT ($\lnot$) and have indegrees of 2, 2 and 1 respectively and unbounded outdegrees. For his purposes, Sima expresses Boolean circuits as Boolean expressions ($f(x_1, x_2, x_3,\dots)$) where input nodes correspond to the variables, and gate nodes correspond to logical operators. This allows him to work with and modify expressions without working expressly with circuits and their diagrams. It should be noted that the number of logical operators in an expression may not be directly correlated to the complexity of the corresponding circuit as gates within circuits are allowed to output to multiple other gates. However, this does not affect Sima's argument as he uses these expressions to argue about the correctness (behavior) of the circuits rather than their complexity.

 A Boolean circuit with no NOT gates is called \emph{monotone}.
 
 A Boolean circuit in which only the input nodes are negated (only the input nodes are inputs to NOT gates) is called a \emph{standard} circuit. Because negations occur only at the input nodes, one can rewrite such a circuit,
 \begin{equation*}
     f(x_1, x_2,\dots,x_n),
 \end{equation*}
 as a circuit with twice as many input nodes but no negations,
 \begin{equation*}
     f(x_1, x_2,\dots,x_n, \lnot x_1, \lnot x_2,\dots,\lnot x_n).
 \end{equation*} 
 In this manner one removes the NOT gates, but for any valid assignment of variables, $(\forall i \in \{1\dots n\})\ [x_i = \var{NOT} (\lnot x_i)]$.

 \subsubsection*{Circuit Complexity}
 The \emph{circuit complexity} of a Boolean function is the size (number of gates) of the smallest Boolean circuit that computes the Boolean function. 

 The \emph{standard circuit complexity} of a Boolean function is the size (number of gates) of the smallest standard circuit that computes the Boolean function. 

 The \emph{monotone circuit complexity} of a Boolean function is the size (number of gates) of the smallest monotone circuit that computes the Boolean function.

 \subsubsection*{The Clique Function and its Monotone Complexity}
 For $1\leq s \leq m$, let $\var {CLIQUE}(m, s)$ be the function of $n = \binom{m}{2}$ variables representing the edges of an undirected graph \textit{G} on \textit{m} vertices, whose value is 1 if and only if \textit{G} contains an \textit{s}-clique.

 The monotone complexity of the clique function is exponential. Razborov initially showed a superpolynomial lower bound. This was improved by Alon and Boppana~\cite{monotonecomplexity} to be exponential.
\section{Summary of Sima's Argument}
Sima builds his argument by attempting to fill in the holes left by Alon and Boppana's analysis of the lower bounds for the monotone complexity of the clique function. Their paper is only able to establish lower bounds for the monotone complexity of the clique function. The nonmonotone complexity of the clique function is thus left unbounded. Sima argues that the nonmonotone complexity of the clique function is in fact greater than or equal to that of the monotone complexity of the clique function.

In order to show this, Sima attempts to transform a nonmonotone circuit for the clique function into a monotone circuit for the clique function without increasing its size beyond a polynomial factor. He first considers standard circuits (as defined previously). He claims that any circuit can be transformed into a standard circuit by at most doubling the number of gates. Because of this, the difference in complexity between standard and non-standard circuits is at most a factor of 2, thus allowing him to consider only standard circuits. From this point he makes his main argument.

He considers the standard Boolean circuit $f(x_1, x_2,\dots,x_n, \lnot x_1, \lnot x_2,\dots,\lnot x_n)$ that computes the clique function, $\var{CLIQUE}(m, s)$. At this point he makes the argument that replacing any one of the negated variables ($\lnot x_i$) with 1 (TRUE) will result in a circuit that computes the same function. To prove this, he first argues that one can ``extract" any negated variable in the circuit, moving it to the front of the formula. This results in a formula of the form\footnote{Throughout this paper, we assume the standard
  operator precedence rules, and in particular that $\lnot$ has higher precedence
  than~$\wedge$, which itself has higher precedence than $\lor$.}
\begin{equation*}
     f=\lnot x_i \wedge \var{TermA} \lor \var{TermB} \lor \var{TermC} \dots,
 \end{equation*}
where $x_i$ is the extracted variable and none of $\var {TermA}$, $\var {TermB}$, $\var {TermC}$\dots include $\lnot x_i$. 

He then uses this extracted form of the formula to argue that replacing $\lnot x_1$ (or some other arbitrary negated variable) with 1 does not change the value of $f$ when $f$ calculates the clique function. The argument is as follows.

 Sima starts with the extracted form of the standard clique formula where $\lnot x_1$ is the extracted variable:
\begin{equation*}
     f=\lnot x_1 \wedge \var{TermA} \lor \var{TermB} \lor \var{TermC} \dots\,\,.
\end{equation*} 
There are then two cases he considers. Sima refers to $\lnot x_1 \wedge \var{TermA}$ as $\var{Term1}$, with $\var{Term1part1}$ referring to $\lnot x_1$ and $\var{Term1part2}$ referring to $\var{TermA}$.

Case 1: $\var{Term1part2}$ has a value of 0.\newline
In this case he argues that because this second part of $\var{Term1}$ has a value of 0, clearly no matter what you set $\lnot x_1$ to, the value of $\var{Term1}$ will be 0. Thus the output of the circuit will not change.

Case 2: $\var{Term1part2}$ has a value of 1.\newline
In this case, he argues that if $\var{Term1part2}$ has a value of 1 then the value of the entire function is also 1. The reasoning he gives is that if $\lnot x_1$ is also 1, then $x_1$ takes the value 0. As such, by the definition of the clique function, the edge corresponding to $x_1$ is disconnected and thus $\lnot x_1$ (and $x_1$) has no contribution to the size of the clique. Thus as long as the value of $\var{Term1part2}$ is 1, the value of the circuit will be 1 no matter what $\lnot x_1$ is. 

The final step in Sima's proof is to argue that since this aforementioned process can be done with any of the negated variables ($\lnot x_1,\ \lnot x_2,\dots, \lnot x_n$), you can (sequentially) replace all the negated variables with the value 1, and the resulting circuit will compute the clique function correctly. Furthermore he states that since this replacement will clearly not increase the complexity of the circuit, and (as you are starting with a standard circuit) the resulting circuit will be monotone, standard circuits for the clique function are no smaller than monotone circuits for the clique function. Since the monotone circuit complexity of the clique function was proven to be exponential, the standard circuit complexity (and thus the complexity overall) of the clique function is also exponential. As such he concludes that $\pisnotnp$.

\section{Critique of Sima's Argument}
On its surface, Sima's argument seems sound. He builds off Alon and Boppana's finding about the monotonic complexity of the clique function by converting a nonmonotone circuit into a monotone one without increasing its complexity. However, the problem lies in his argument about the conversion process. From here let's follow his argument more closely. 

He starts with a(n) (arbitrary) standard circuit that computes the clique function. This is then expressed in the form $f = f(x_1, x_2,\dots,x_n, \lnot x_1, \lnot x_2,\dots,\lnot x_n)$. His first claim is that any of the negated variables ($\lnot x_i$) can be extracted, putting the circuit into the form $f=\lnot x_i \wedge \var{TermA} \lor \var{TermB} \lor \var{TermC} \dots$, where each of  the terms $\var{TermA}$, $\var{TermB}$,\dots does not contain $\lnot x_i$. This is in fact trivial. By first expanding the Boolean formula into Sum of Products form and then reorganizing and factoring out $\lnot x_i$, this is easily achievable.

He then considers extracting one of the negated variables (using $\lnot x_1$ as his example, but extending the argument to any negated variable) and makes arguments for two cases concerning the value of $\var{Term1}$ (again where $\var{Term1} = \lnot x_1 \wedge \var{TermA}$, $\var{Term1part1} = \lnot x_1$, and $\var{Term1part2} = \var{TermA}$). 

In the first case he considers when $\var{Term1part2}$ has a value of 0. His argument in this case is sound. When $\var{Term1part2}$ has a value of 0, the contribution from $\var{Term1}$ to the overall function will be 0 no matter the value of $\lnot x_1$.

However, in his second case he runs into trouble. When the value of $\var{Term1part2}$ is 1, he again tries to present an argument for why setting the value of $\lnot x_1$ to 1 will not change the value of the function. But here he misunderstands what he is actually arguing and neglects to fully comprehend the connection between $x_1$ and $\lnot x_1$. His initial argument is true. If $\var{Term1part1}$ is 1, then clearly the value of $f$ will be 1. Due to the clique function being monotone, this does in fact mean that the $x_1$ edge has no contribution to the value of the clique function as adding an edge to the graph (changing the value of $\lnot x_1$ from 1 to 0 and vice versa for $x_1$) will not cause a clique in the graph to disappear. As such any assignment of $\lnot x_1$ will result in the same value of the function. A more formal description of what Sima proves follows below.
\begin{definition}
Let $A$ be an assignment of the Boolean variables $x_1, x_2, x_3,\dots,x_n$. Define $A'$ to be the same assignment as $A$ except with the value of $x_1$ reversed (changed from 0 to 1 or vice versa).
\end{definition}
\begin{claim}
If both $\var{Term1part1}$ and $\var{Term1part2}$ evaluate to 1 given an assignment $A$, then both $f(A) = 1$ and $f(A') = 1$. The same is true swapping $A$ and $A'$, as $A$ = $A''$
\end{claim}
\noindent
By his argument Sima is able to prove the claim above: that if a variable assignment $A$, or its corresponding assignment $A'$, in which the assigned value of $x_1$ is reversed, causes the value of $\var{Term1}$ to be 1, then BOTH $f(A) = 1$ and $f(A') = 1$. However, Sima mistakenly assumes that all assignments where $\var{Term1part2} = 1$ fall into the form of $A$ or $A'$. This appears, at first, to be true as $\var{Term1part1} = \lnot x_1$. So by flipping the value of $\lnot x_1$ (i.e. changing from $A$ to $A'$) you can always make $\var{Term1}$ evaluate to 1. However, this ignores the connection between $x_1$ and $\lnot x_1$. Because $x_1$ must be equal to the negation of $\lnot x_1$ in any valid assignment, it is possible for there to be variable assignments $B$ for which the value of $\var{Term1part2}$ is 1, while neither $B$ nor $B'$ result in both $\var{Term1part1}$ AND $\var{Term1part2}$ having a value of 1. We will describe one such case below.

\subsection{An Illustrative Example}
Consider the following example:
Let $f(x_1, x_2,\dots,x_n)$ be a monotone circuit that computes the clique function, $\var{CLIQUE}(m, s)$, for some $s\geq 3$.
Now append to the circuit (via the logical OR operator) the term
\begin{equation*}
    \lnot x_1 \wedge x_1.
\end{equation*}
The resulting circuit $f'$ is now
\begin{equation*}\label{fp}
    f'(x_1, x_2,\dots,x_n) = x_1 \wedge \lnot x_1 \lor f(x_1, x_2,\dots,x_n),
\end{equation*}
or in standard form
\begin{equation*}
    f'(x_1, x_2,\dots,x_n, \lnot x_1) = x_1 \wedge \lnot x_1 \lor f(x_1, x_2,\dots,x_n).
\end{equation*}
Since the appended term is NEVER satisfied and is adjoined to $f$ via an OR operator, the resulting $f'$ will have the same behavior as $f$ and will also calculate $\var{CLIQUE}(m,s)$ correctly. Now following Sima's process and extracting the negated $\lnot x_1$, the result is
\begin{equation*}
    f'(x_1, x_2,\dots,x_n, \lnot x_1) = \lnot x_1 \wedge (x_1) \lor \var{TermB} \lor \var{TermC}\dots,
\end{equation*}
where $\var{TermB}$, $\var{TermC}$,\dots are terms (without negation) containing any of the variables except $\lnot x_1$. Note that since $f$ is a monotone circuit, the only negated variable will be the $\lnot x_1$ that was just introduced. What Sima refers to as $\var{Term1}$ in this case is $\lnot x_1 \wedge x_1$, with $\var{Term1part1}$ being $\lnot x_1$ and $\var{Term1part2}$ being $x_1$. Now, per Sima's algorithm, set $\lnot x_1$ to 1, resulting in the following circuit $f''$:
\begin{eqnarray*}
    f''(x_1, x_2,\dots,x_n, 1) &=& 1 \wedge (x_1) \lor TermB\dots\\& =& x_1 \lor TermB\dots\,\,.
\end{eqnarray*}
From here it is clear to see that as long as $x_1$ has a value of 1, the value of $f''$ will be 1. Looking at the equation from another perspective, even if ONLY $x_1$ has a value of 1 and all the other variables are set to 0, the circuit will still output 1. However, bringing this back to our definition of the clique function, if there is a graph with only one edge, it is impossible to have a clique of size anything greater than 2. Thus this new $f''$ clearly does not compute the same function as $f$.

The reason Sima's argument fails is that the set of all variable assignments $A$ for which both $\var{Term1part1}$ and $\var{Term1part2}$ evaluate to 1, and their corresponding assignments $A'$, does not equal the set of all assignments S for which $\var{Term1part2}$ evaluates to be 1. This is stated more formally below. Let $\var{Term1part2}(A)$ be the value of $\var{Term1part2}$ given the assignment $A$ and similarly for $\var{Term1part1}(A)$. Then
\begin{align*}
    \{A \mid \var{Term1part2}(A) = 1,\var{Term1part1}(A) = 1\}\ \cup\ &\\
    \{A \mid \var{Term1part2}(A') = 1,\var{Term1part1}(A') = 1\} &\neq \{A \mid  \var{Term1part2}(A) = 1\}.
\end{align*}
\noindent
In fact, there are no valid assignments $A$ (and thus no corresponding $A'$) for which both $\var{Term1part1}$ and $\var{Term1part2}$, in the example presented in this section, evaluate to 1. However, any assignment in which $x_1$ is 1 causes $\var{Term1part2}$ to evaluate to 1. Because $x_1$ and $\lnot x_1$, although treated as different variables in the standard formula, must be negations of each other in any valid assignment, it is possible to create cases that Sima's argument fails to cover such as the one presented above.

By failing to consider this connection between $\lnot x_1$ and $x_1$, Sima makes an intuitive generalization that, although on the surface may seem reasonable, leaves loopholes that can change the behavior of the function. The above example illustrates the error in his logic and provides a specific counterexample to his process.

\section{Conclusion}
Sima's argument attempts to build off of the findings of Alon and Boppana~\cite{monotonecomplexity} in order to extend the exponential lower bound for the monotone circuit complexity of the clique function, to the nonmonotone circuit complexity of the clique function. He describes a clever attempt to convert standard (nonmonotone) circuits into monotone circuits and attempts to prove that this conversion holds in the case of the clique function. However, in doing so he makes an unfounded generalization in his argument. This results in specific cases that can be exploited by using the connection between variables and their negations. 

By describing his flaw and presenting a counterexample to his process, we have demonstrated that his method is not satisfactory. It is possible that through using a more specific description of what the minimal nonmonotone circuit of the clique function looks like, one could sidestep the problems that we have described in this paper and establish an exponential lower bound on the nonmonotone complexity of the clique function. In fact, in 2005 Amano and Maruoka were able to show that the lower bound for the complexity of nonmonotone circuits for the clique function with at most $1/6\log(\log(n))$ negation gates is in fact superpolynomial~\cite{smallnegs}. However, until such time as we have a better understanding of the clique function and its properties, a proof such as presented in Sima's paper is not possible.  
\bibliography{references}
\end{document}